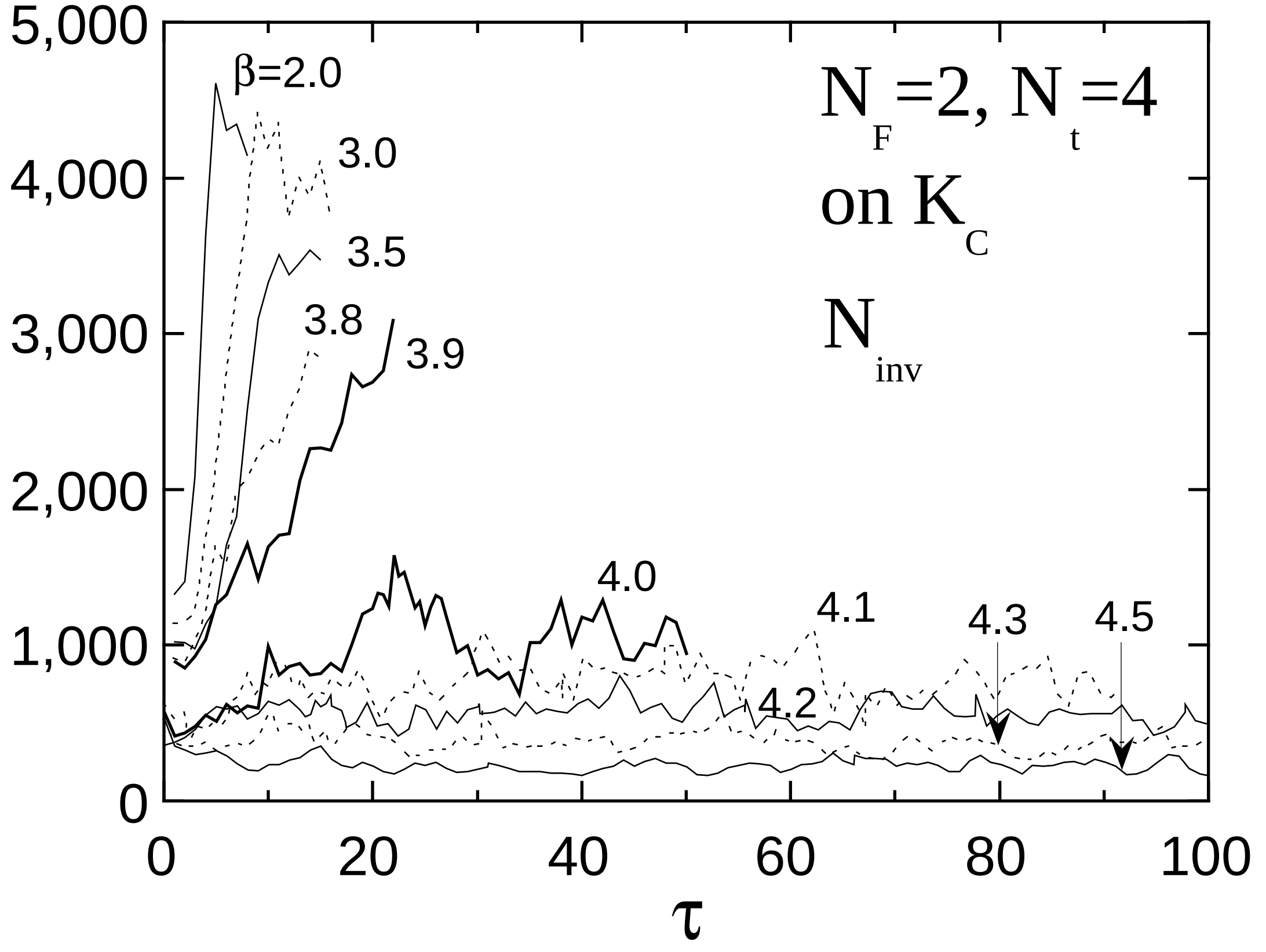

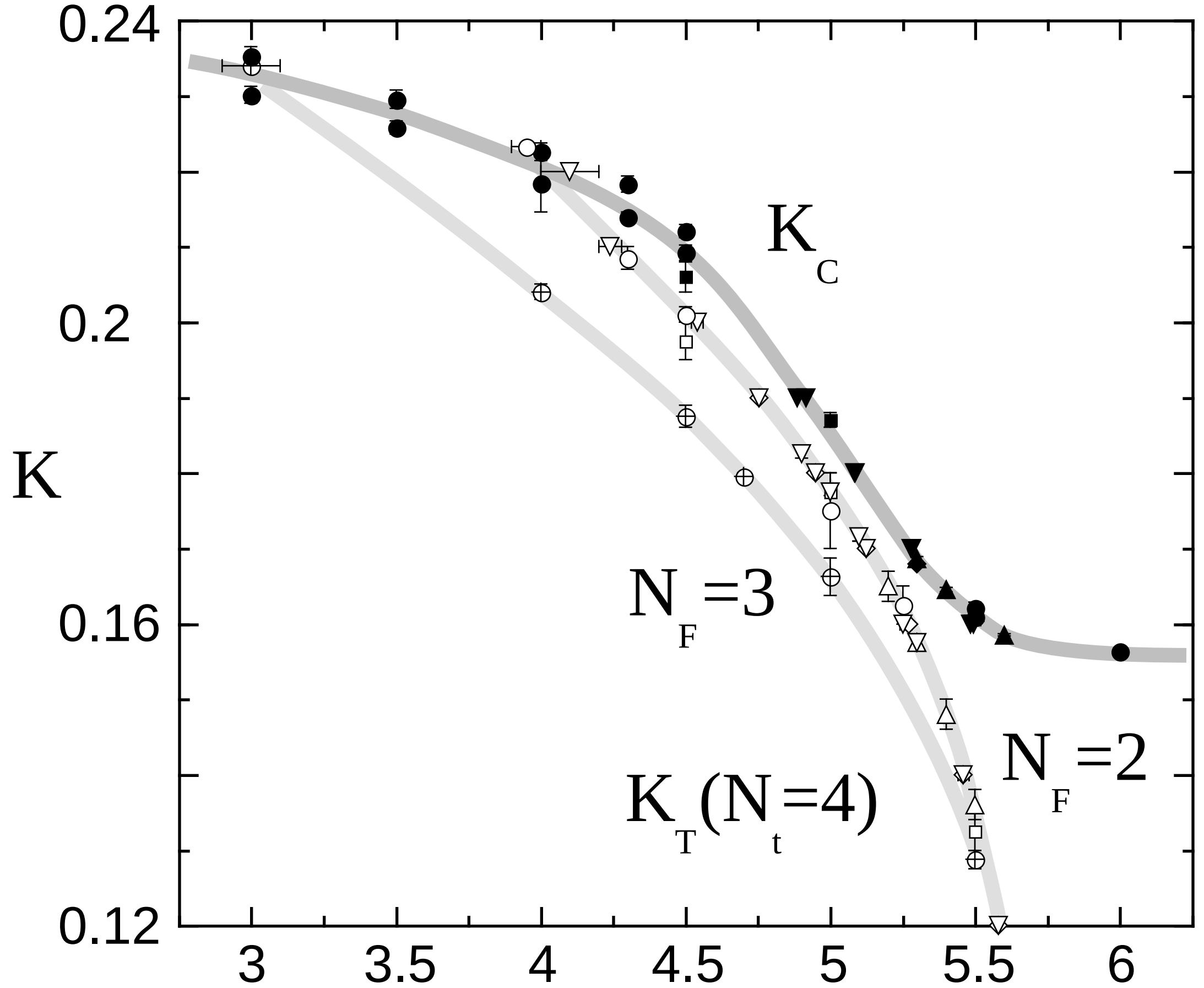

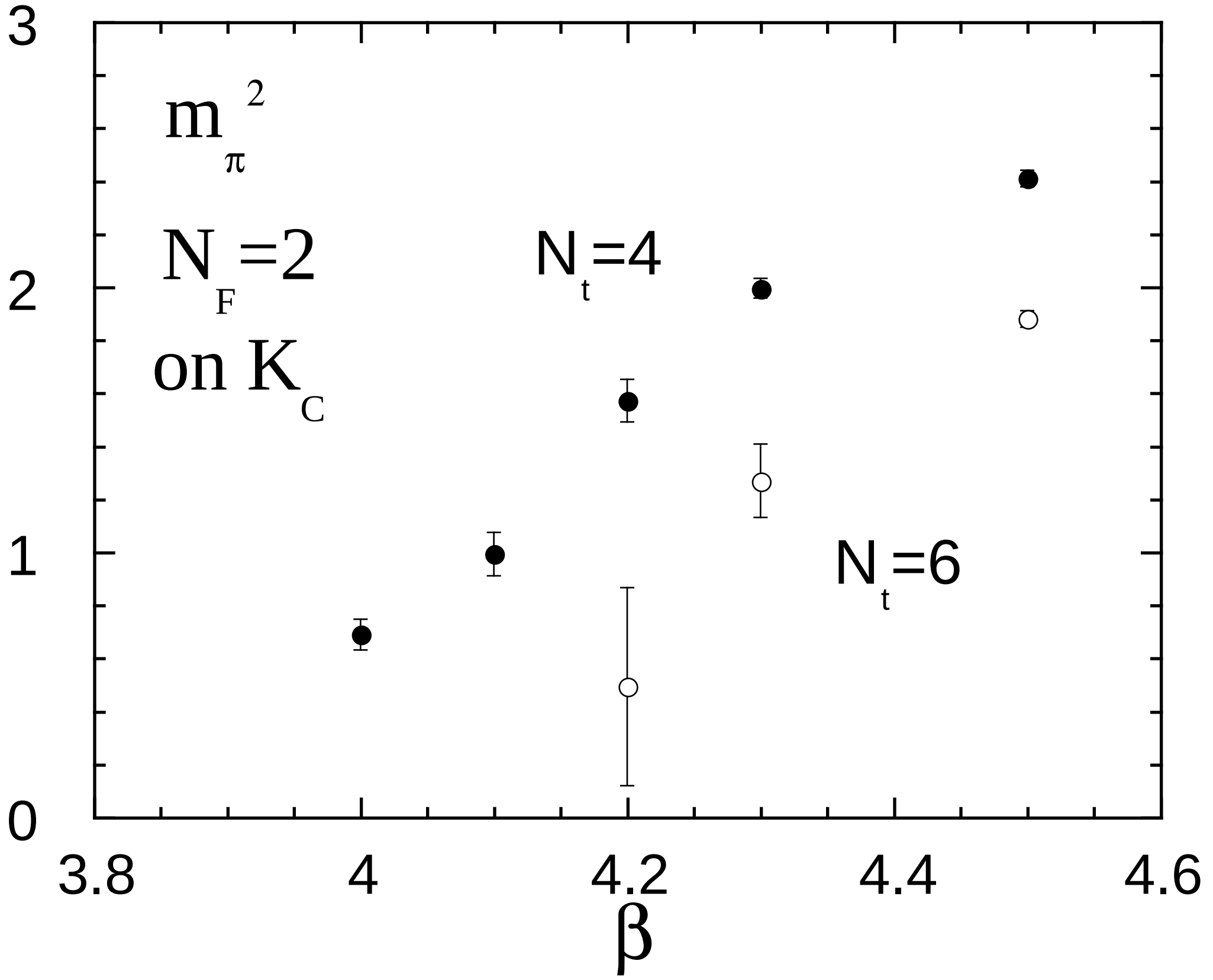

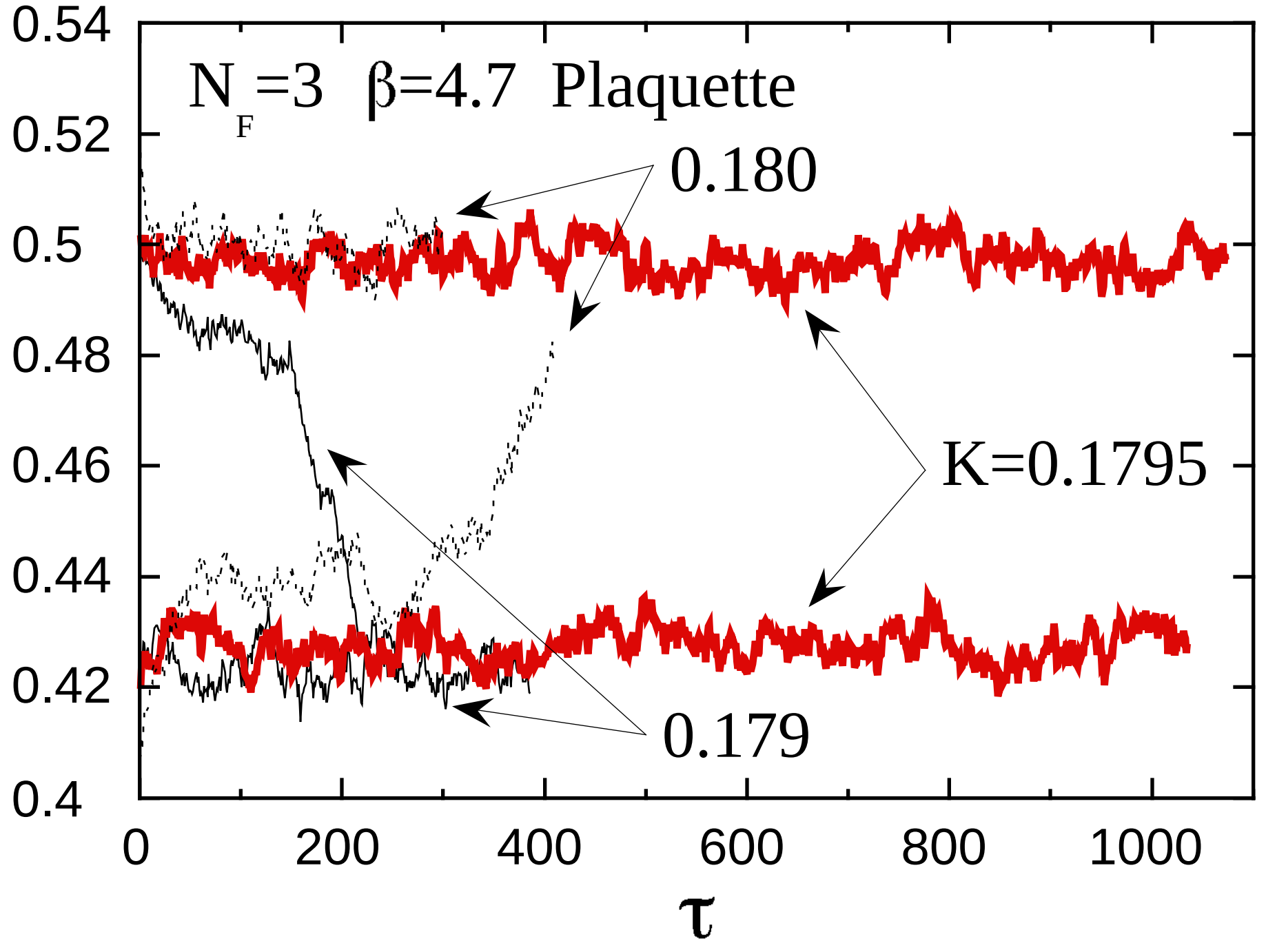

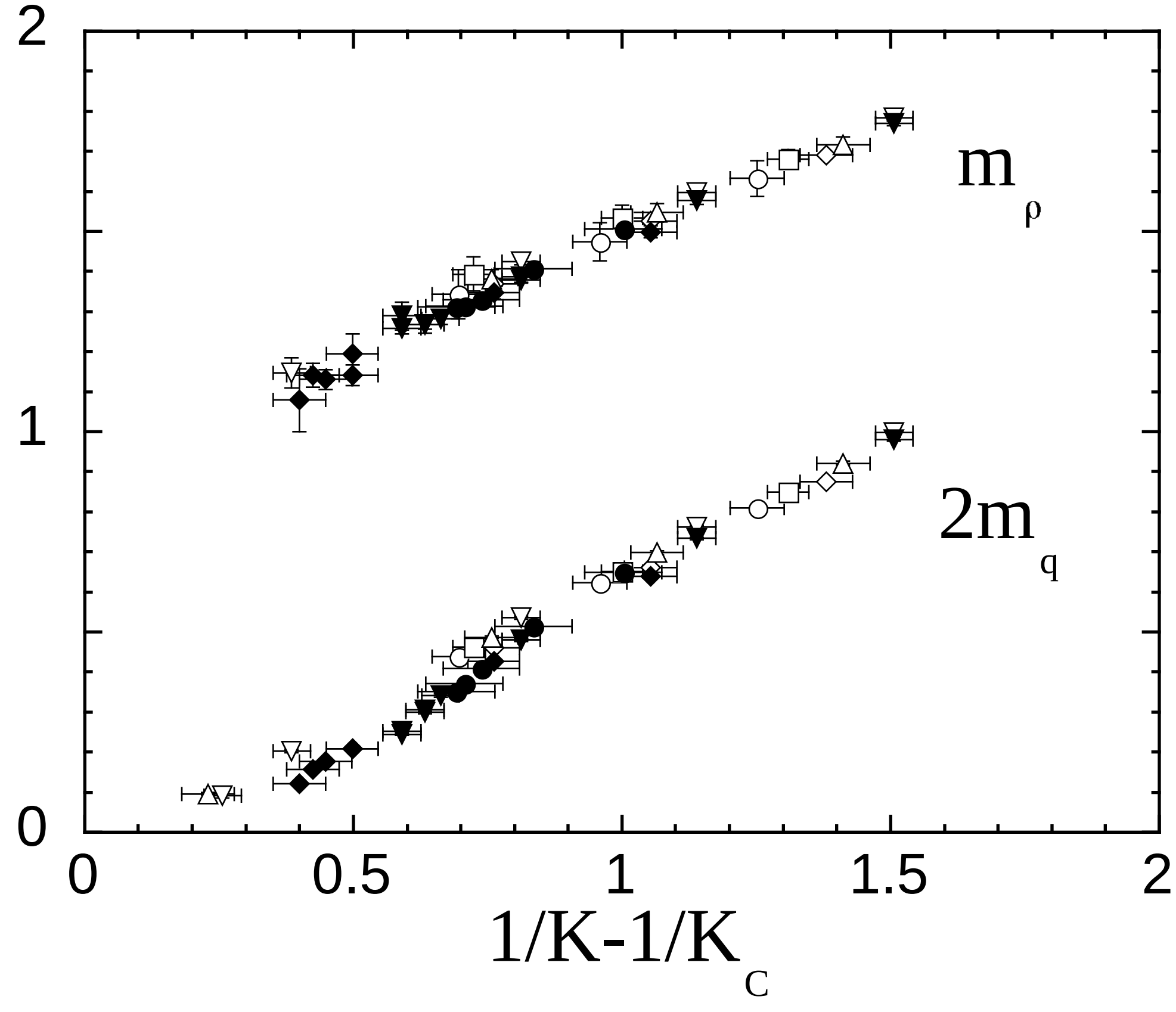

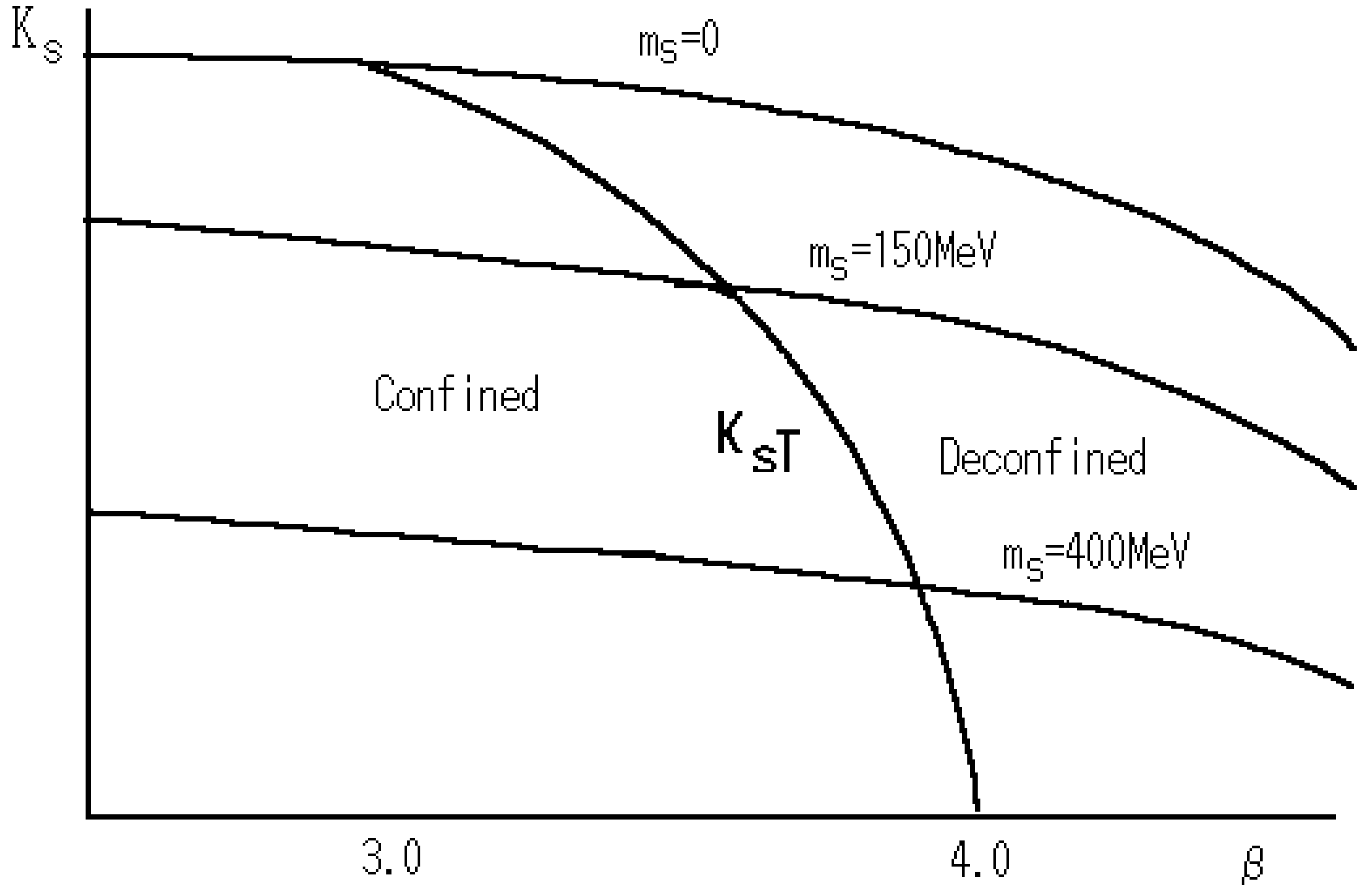

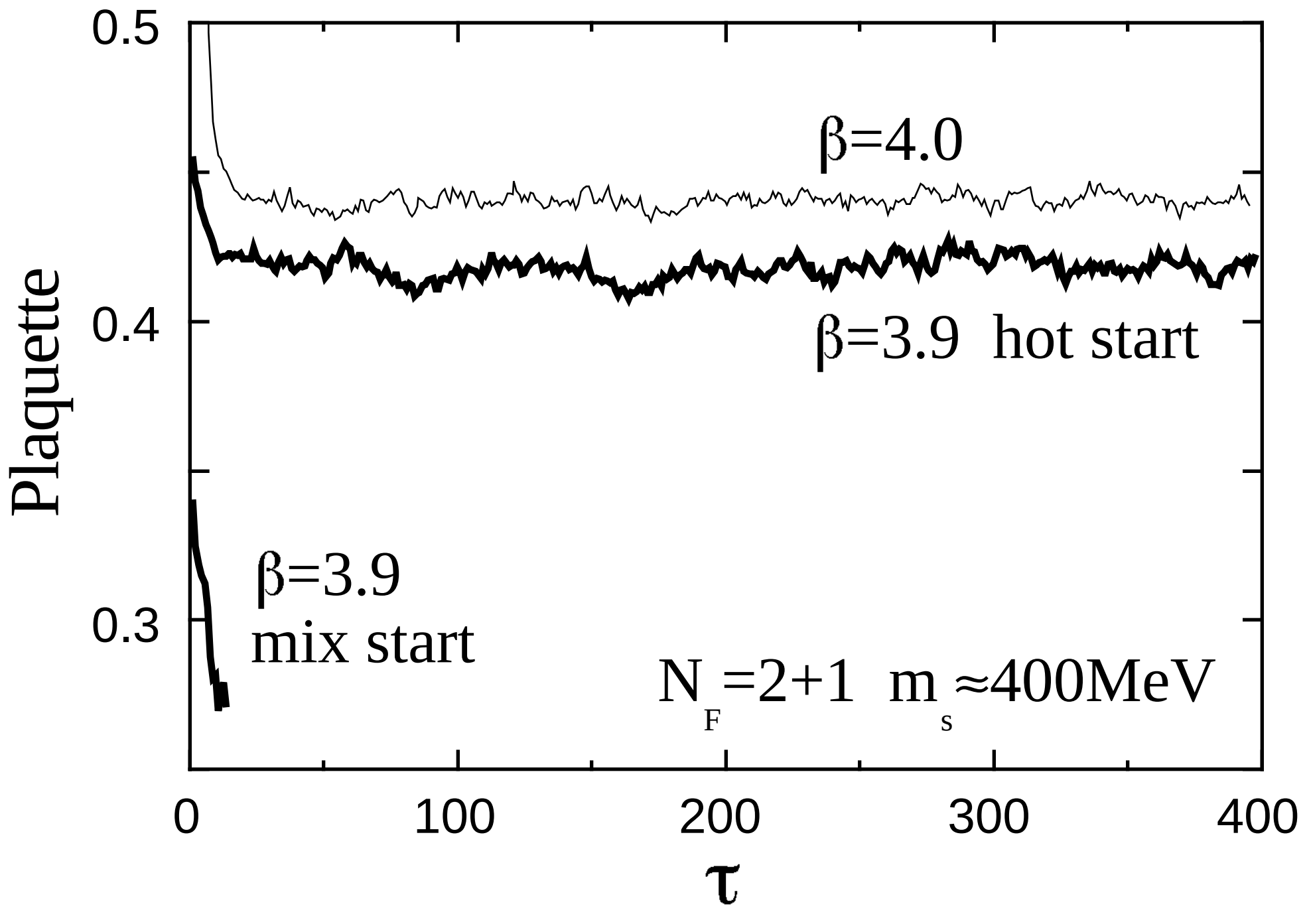

# Finite Temperature QCD with Wilson Fermions

S.Sakai†
Faculty of Education, Yamagata University, Yamagata 990, Japan

**Abstract**

The nature of finite temperature transition in QCD is studied on a lattice with Wilson fermion. For massless quarks, the transition is smooth for $N_F = 2$, while it is of first order for $N_F = 3$ and 6. These results are consistent with the prediction based on universality. For the massive degenerate $N_F = 3$ case, clear two states signals are observed for $m_q \lesssim 140 MeV$. From the study of $N_F = 2+1$ case, namely two massless u-, d-quark and a s-quark with light mass, our results suggest that the finite temperature transition in the real world is first order.

## 1 Introduction and The Chiral Properties of Wilson Action

The nature of finite temperature transition in QCD is a fundamental information for the study of quark-gluon plasma formation in high energy heavy ion collisions, and the evolution of the universe at the very early stage.

In this article we discuss on the point " what is the nature of finite temperature phase transition with realistic quark masses". Quarks with much heavier masses than the transition temperature will decouple from the dynamics of transition. Because the physical strange quark mass is the same order of magnitude as the temperature of the transition; $100 \sim 200 MeV$ ,we should take into account the dynamics of the u-, d- and s-quark properly.

†in collaboration with Y.Iwasaki[a,b], K.Kanaya[a,b],S.Kaya[a], and T.Yoshie[a,b]
[a]Institute of Physics, University of Tsukuba, Ibaraki 305,Japan
[b]Center for Computational Physics, University of Tsukuba, Ibaraki 305,Japan

In this article, we use Wilson fermion action. With this action quarks in any number of flavor($N_F$) are treated with a local action. While the chiral symmetry is explicitly broken by the Wilson term even for vanishing bare quark masses. However chiral properties have been carefully investigated by the following axial vector Ward identity[1].

$$2m_q < \pi|P_5|0 >= -m_\pi < \pi|A_4|0 > +O(a) \tag{1}$$

where $A_4$ is the fourth component of axial vector current , and $P_5$ is pseudoscalar respectively.

From numerical studies it is found that in the confined phase, $m_q$ defined by Eq.(1) is proportional to $m_\pi^2$. It is further found that in the large $\beta(\beta \geq 5.5$ for $N_F = 2)$ region, the values of quark mass are independent of whether it is calculated in confined phase or deconfined phase[2],[3],[4]. In the small $\beta$ region however, $m_q$ in the deconfined phase does not agree with that in the confined phase and shows peculiar behavior if plotted as a function of $1/K$, which may be due to $O(a)$ lattice artifacts[4],[5].

Chiral limit($K_c(\beta)$) is defined by extrapolating either $m_\pi^2$ or $m_q$ to zero in the confined region for $\beta \leq 5.5$, while for larger $\beta$ region it can also be defined directly by the point $m_q = 0$.

We further define $K_t$, which is the hopping parameter at which the finite temperature phase transition occurs. It is a function of temporal size of lattice $N_t$, number of flavor $N_F$ and $\beta$. As first noticed by Fukugita *et al.*[6], whether the $K_t$ line crosses $K_c$ line or not was not trivial. This question can be answered if we know whether the point $\beta = 0$ and $K_c = 0.25$ is in the confined phase or deconfined phase when number of flavor is changed. The confinement property at $\beta = 0$ has been studied by Iwasaki *et al.*[7]. It has been concluded that for $N_F \geq 7$ the point $\beta = 0$ and $K = K_c$ is in the deconfined phase even in zero temperature(we have confirmed deconfinement up to $N_t = 18$) and for $N_F \leq 6$ it is in the confined phase for $N_t \geq 4$ .

In this article we present the results of simulations on $8^2 \times 10 \times 4$, $12^3 \times 4$ and $12^3 \times 6$ lattices. In the calculation of hadron mass, the size of third direction is duplicated. For the further details of simulations see[8] and [9]. In section 2, we show that the nature of chiral transition is continuous for $N_F = 2$ and first order for $N_F = 3$ and 6. In section 3 the nature of the transition with massive quarks is discussed for $N_F = 3$. In the case of degenerate $m_q$, the first order transition is observed for $m_q \lesssim 140 MeV$. For $N_F = 2+1$ case, namely $m_u = m_d \simeq 0$ and $m_s \neq 0$, two states signals are observed both for $m_s \sim 150 MeV$ and $m_s \sim 400 MeV$. Therefore our results indicate first order transition for the real world. The section 5 is devoted to conclusions and discussions.

# 2 Chiral Phase Transition of Lattice QCD

In this section, we study the nature of phase transition at $m_q = 0$, namely at the point where $K_c$ line and $K_t$ line crosses. We denote corresponding $\beta$ as $\beta_{ct}$. For this study we take a strategy to perform the simulation on $K_c(\beta)$ line starting from the deconfined region and decrease $\beta$ until $\beta_{ct}$. We call this method as "on $K_c$" simulation method.

It has been shown that in the confined phase, the zero eigenvalues appear in the quark matrix around $K_c$, while in the deconfined phase they are not small[7]. Therefore the number of iteration for quark matrix inversion $N_{inv}$ is a good indicator to discriminate the difference of phase on $K_c$. In addition to $N_{inv}$, the measurements of Polyakov loops, Wilson loops and hadron screening masses are used to discriminate the differences of phases.

## 2.1 $N_F = 2$

Time history of $N_{inv}$ for $N_t = 4$ is shown in Fig. 1 . Up to $\beta = 4.0$ the value of $N_{inv}$ does not exceed the order of several hundreds, but for $\beta \leq 3.9$ it increases rapidly with trajectory. Time histories of plaquette, Polyakov loop, $m_\pi^2$ etc also confirm that the the states for $\beta \leq 3.9$ are going into the confined phase. Therefore we conclude $\beta_{ct} \simeq 3.9 - 4.0$ for $N_t = 4$. Similar study for $N_t = 6$ results in $\beta_{ct} \simeq 4.0 - 4.2$. The $K_t(\beta)$ line for $N_t = 4$ and $K_c(\beta)$ lines are shown in Fig. 2[6],[3],[10],[11]. The crossing points $\beta_{ct}$ obtained by "on $K_c$" simulation method are consistent with the points determined by the extrapolation of the $K_t$ line toward $K_c$, both for $N_T = 4$ and 6.

The $m_\pi^2$ on $K_c$ in the deconfined phase, are shown in Fig 3. They are decreasing gradually and at the region of $\beta_{ct}$, they are consistent with zero. Around $\beta_{ct}$ we could not observe two states signal. From these results we conclude that the phase transition in this case is continuous.

### 2.2 $N_F = 3, 6$

The same method as in the case of $N_F = 2$ is applied for the study of $N_F = 3$ and $N_t = 4$. We observe clear two states signal at $\beta = 3.0$; if a initial condition is deconfined states, $N_{inv}$ is stable around several hundreds and system stays in the deconfined phase, while if we start from a mixed states, $N_{inv}$ increases with trajectory. Similarly the hadron screening mass, quark mass, plaquette and Polyakov loop also suggest two states signal at $\beta = 3.0$[8].

The situation for $N_F = 6$ is completely same as the case of $N_F = 3$ except that two-states signal is observed at $\beta = 0.3$.

From these results we concludes that the nature of chiral transition for $N_F = 2$ is continuous while for $N_F = 3$ and 6 , first order. These results are consistent with those from staggerd quark[12], and the prediction based on universality arguments[13].

## 3 Finite Temperature Phase Transition with Massive Quark

### 3.1 Degenerate $N_F = 3$ Case

Before proceeding to the realistic non-degenerate quark mass case, we study degenerate $N_F = 3$ case. In this case the simulation is made on $K_t$ line for $N_F = 3$ in Fig. 2. We observe clear two states signal at $\beta = 4, 0, 4, 5$ and 4.7(Fig. 4). However no two states signal has been observed at $\beta = 5.0$ and 5.5. At $\beta = 5.0$ we have made more fine tuning of $K$ than in the case of $\beta = 4.7$[9].

Then the critical quark mass $m_q^{crit}$ up to which the clear two states signal is observed is bounded from below by the $m_q$ at $\beta = 4.7$ and $K = 0.1795$. That is $m_q^{crit} a \geq 0.175$.

The lattice spacing around this $\beta$ region is determined by the $\rho$ meson spectroscopy. In Fig. 5 the $m_\rho a$ and $m_q a$ are plotted as a function of $1/K - 1/K_c$. They are almost independent of $N_F$, $N_t$ and $\beta$. From $m_\rho a$ at $K_c$ and physical $\rho$ meson mass($m_\rho = 770MeV$), we get $a^{-1} \simeq 0.8GeV$. Then $m_q^{crit}$ in physical unit becomes $\gtrsim 140MeV$.

### 3.2 $N_F = 2 + 1$

We proceed to the study of realistic quark mass case: $m_u = m_d \simeq 0$ and $m_s \neq 0$. As the $m_q^{crit}$ for degenerate $N_F = 3$ case discussed in the previous section is almost same order of magnitude as the physical strange quark mass, the nature of transition will depends on $m_s$ sensitively.

In this case, confined and deconfined phases are separated by the finite temperature transition line of strange quark $K_{sT}$, which is shown schematically in Fig. 6. Along $K_{sT}$ line, the nature of transition changes from first order at $\beta \simeq 3.0$ to continuous when $\beta$ is increased. We want to know whether the critical quark mass of s-quark $m_s^{crit}$ at which the nature of transition changes from first order to continuous is lighter than physical strange quark mass or not.

For this study we use the similar method as "on $K_c$" simulation method; namely keeping $K_u = K_d = K_c$, and $K_s$ to the value corresponding to of $m_s \sim 150MeV$ or $400MeV$, and decrease $\beta$ in the deconfining region, until we hit $K_{sT}$ line. The $K_s$ for $m_s \sim 150MeV$ and $400MeV$ are obtained from Fig. 5[9] and schematically shown in Fig. 6.

In Fig. 7 we show time history of plaquette at the crossing point for $m_s \sim 400MeV$. Clear two states signal is observed at $\beta = 3.9$, both on $8^2 \times 10 \times 4$ and $12^3 \times 4$ lattices. The mesurements of plauette, Polyakov loop and $m_\pi^2$ etc also indicate two states signal[9]. For $m_s \sim 150MeV$ too, two states signal is observed at $\beta = 3.5$.

## 4 Conclusions and Discussions

Our results implies that chiral transition is continuous for $N_F = 2$ (both for $N_t = 4$ and 6), and for $N_F = 3$ and 6 it is first order. These results are consistent with those of staggerd quark[12], and the prediction based on universality[13]. For massive degenerate $N_F = 3$ case, our result imply $m_q^{crit} \gtrsim 140MeV$ ,and for $N_F = 2 + 1$ $m_s^{crit} \gtrsim 400MeV$.

These critical quark masses are much heavier than those of staggerd quark[14],[15]. Columbia group reported that first order transition is not observed at $m_u a = m_d a = 0.025$, $m_s a \simeq 0.1$. Using $a^{-1} \simeq 0.5GeV$ for staggerd quark around this $\beta$ region[9],[16] ,the quark masses in the physical unit are $m_u = m_d \simeq 12MeV$ and $m_s \simeq 50MeV$ respectively. One of the pssibilities for the discrepancy will be that $\beta$ in the both simulations are far from continuum limit. To conclude the nature of transition in the real world, further studies with larger $\beta$ are necessary, but our result suggests a possibility for the first order transition in QCD in the real world.

Simulations are performed with HITAC S820/80 at KEK, and with VPP-500/30 and QCDPAX at University of Tsukuba. We thank members of KEK for their hospitality and strong support. We also thank other menmers of QCDPAX collaboration for their support. This work is in part supported by the Grant-in-Aid of Ministry of Education, Science and Culture(No.06NP0601).

=13cm Fig1.ps

Figure 1: Time history of $N_{inv}$ on $K_c$ for $N_F = 2$ on $8^2 \times 10 \times 4$ lattice.

=13cm Fig2.ps

Figure 2: Phase diagram for $N_F = 2$ and 3. Filled symbols are for $K_c(m_\pi^2)$ and $K_c(m_q)$. Open symbols are for $K_t(N_t = 4)$ for $N_F = 2$ while open circles with cross for $N_F = 3$. Circles are our data. Lines are for guiding eyes.

=13cm Fig3.ps

Figure 3: $m_\pi^2 a^2$ on the $K_c$ line for $N_t = 4$ and 6 lattice. The spacial sizes are $8^2 \times 20$ and $12^2 \times 24$ respectively, where original lattice are duplicated in the 3-rd direction.

=10.3cm Fig4.ps

Figure 4: Time history of the plaquette for $N_F = 3$ at $\beta = 4.7$ on a $12^3 \times 4$ lattice.

=12cm Fig5.ps

Figure 5: $m_\rho a$ and $2m_q a$ in the confined phase. Open symbols are for $N_F = 2$, $\beta = 3.0$, 3.5, 4.0, 4.3, and 4.5 on an $8^2 \times 10 \times 4$ lattice. Filled symbols are for $N_F = 3$, $\beta = 4.0$, 4.5 and 4.7 on $8^2 \times 10 \times 4$ and $12^3 \times 4$ lattices. $K_c$ for $N_F = 2$ is used.

=12cm Fig6.ps

Figure 6: Phase diagram for $N_F = 2{+}1$ as a function of $K_s$ at $K_u = K_d = K_c$. $\beta_{ct}$ for $N_F = 2$ is $\sim 4.0$ and for $N_F = 3$ is $\sim 3.0$

=12cm Fig7.ps

Figure 7: Time history of the plaquette for $m_s \sim 400$ MeV on a $12^3 \times 4$ lattice.